\documentclass[runningheads]{llncs}
\usepackage[T1]{fontenc}
\usepackage{graphicx}

\usepackage{makecell}
\usepackage{multirow}
\usepackage{amsmath}
\usepackage{mathrsfs}
\usepackage[bookmarks=true]{hyperref}
\usepackage{lipsum}
\usepackage{xcolor}
\usepackage{bbding}

\definecolor{gray}{RGB}{128,128,128}
\SetLipsumParListSurrounders{\begingroup\color{gray}}{\endgroup}
\begin{document}
\title{Boundary-Aware Network for Kidney Parsing}
%
%
\author{
Shishuai Hu
\and
Yiwen Ye
\and
Zehui Liao
\and
Yong Xia\Envelope
} 
\authorrunning{S. Hu et al.}
\institute{National Engineering Laboratory for Integrated Aero-Space-Ground-Ocean Big Data Application Technology, School of Computer Science and Engineering, Northwestern Polytechnical University, Xi’an 710072, China \\
\email{yxia@nwpu.edu.cn}}
\maketitle              
%

\begin{abstract}
Kidney structures segmentation is a crucial yet challenging task in the computer-aided diagnosis of surgery-based renal cancer.
Although numerous deep learning models have achieved remarkable success in many medical image segmentation tasks, accurate segmentation of kidney structures on computed tomography angiography (CTA) images remains challenging, due to the variable sizes of kidney tumors and the ambiguous boundaries between kidney structures and their surroundings.
In this paper, we propose a boundary-aware network (BA-Net) to segment kidneys, kidney tumors, arteries, and veins on CTA scans.
This model contains a shared encoder, a boundary decoder, and a segmentation decoder.
The multi-scale deep supervision strategy is adopted on both decoders, which can alleviate the issues caused by variable tumor sizes.
The boundary probability maps produced by the boundary decoder at each scale are used as attention to enhance the segmentation feature maps.
We evaluated the BA-Net on the Kidney PArsing (KiPA) Challenge dataset and achieved an average Dice score of 
89.65$\%$ for kidney structure segmentation on CTA scans using 4-fold cross-validation.
The results demonstrate the effectiveness of the BA-Net.

\keywords{Boundary-aware network  \and Medical image segmentation \and Kidney parsing}
\end{abstract}

\section{Introduction}
\begin{figure}[t]
  \centering
  \includegraphics[width=0.9\textwidth]{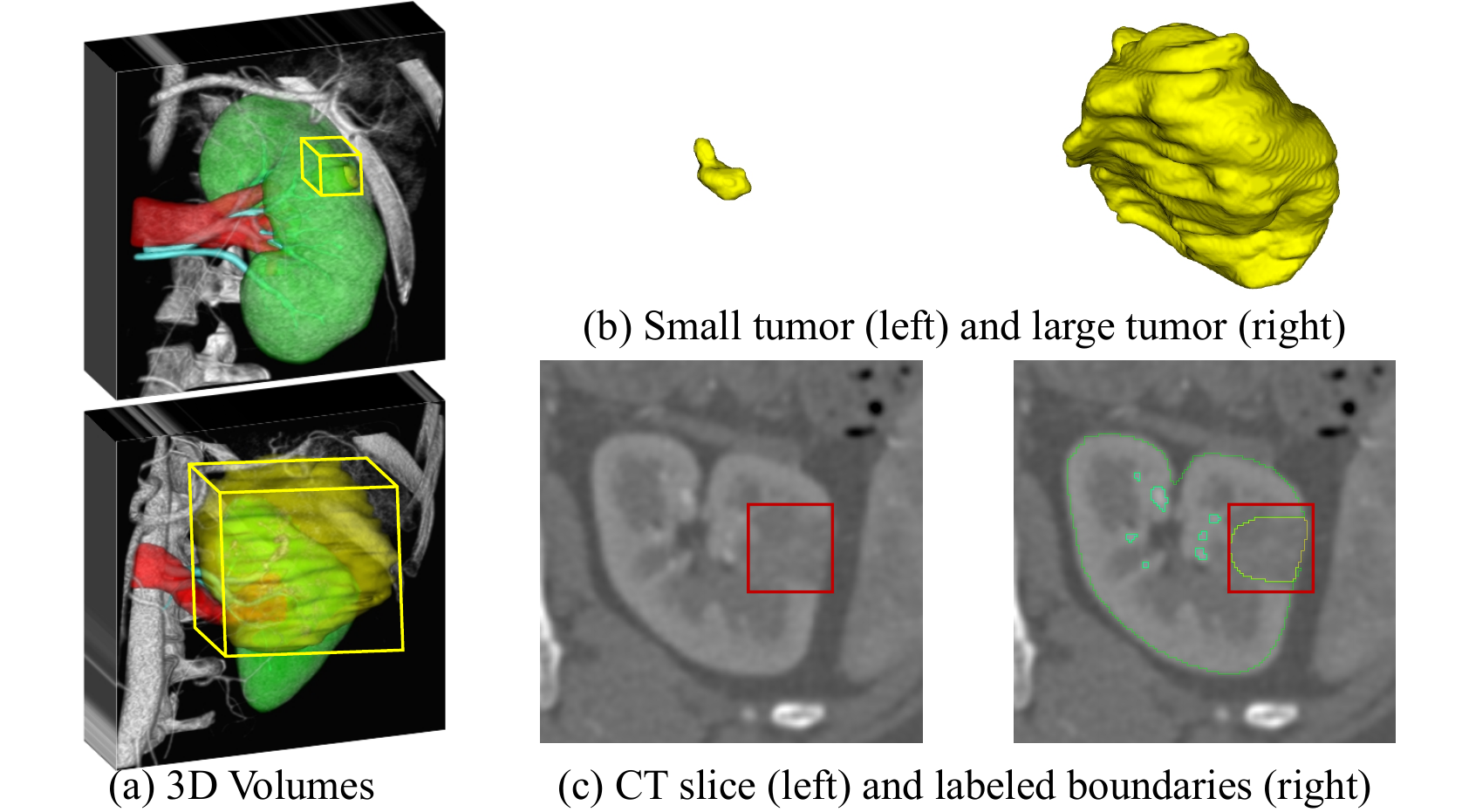}
  \caption{Illustrations of (a) 3D kidney regions of interests (ROIs), (b) imbalanced tumor volumes, and (c) ambiguous boundaries between kidney structures and their surroundings.}
  \label{fig:challenges}
\end{figure}

Accurate kidney-related structures segmentation using computed tomography angiography (CTA) images provides crucial information such as the interrelations among vessels and tumors as well as individual positions and shapes in the standardized space, playing an essential role in computer-aided diagnosis applications such as renal disease diagnosis and surgery planning~\cite{shao2011laparoscopic,shao2012precise,he2020dense,he2021meta}.
Since manual segmentation of kidney structures is time-consuming and requires high concentration and expertise, automated segmentation methods are highly demanded to accelerate this process.
However, this task remains challenging due to two reasons: 
(1) the size of different subtypes of kidney tumors, $e.g.$, papillary tumor and clear cell carcinoma, vary significantly in the volumes, as shown in~\figurename{~\ref{fig:challenges}} (b); and
(2) the contrast between kidney structures and their anatomical surroundings is particularly low, resulting in the blurry and ambiguous boundary, as shown in~\figurename{~\ref{fig:challenges}} (c).

With the recent success of deep convolutional neural networks (DCNN) in many vision tasks, a number of DCNN-based methods have been proposed for kidney-related structures segmentation.
Milletari \textit{et al.}~\cite{milletari2016v} proposed a simple but efficient V-shape network and a Dice loss for volumetric medical image segmentation.
Based on that, Isensee \textit{et al.}~\cite{isensee_nnu-net_2021} integrated the data attributes and network design using empirical knowledge and achieved promising performance on the kidney tumor segmentation task~\cite{heller2021state}.
Following the same philosophy, Peng \textit{et al.}~\cite{Peng_2022_CVPR} and Huang \textit{et al.}~\cite{huang2022adwunet} searched networks from data directly using NAS and further enhanced the accuracy of abdominal organ segmentation.
Although achieved promising performance, these methods~\cite{milletari2016v,isensee_nnu-net_2021,Peng_2022_CVPR,huang2022adwunet} ignore the boundary information, which is essential for kidney structure segmentation.
Compared to the largely imbalanced volume proportions of different tumors, the boundary ($a.k.a.$ surface in 3D) proportions of different tumors are less sensitive to the variable sizes.
Therefore, a lot of boundary-involved segmentation methods have been recently proposed for medical image segmentation.
Kervadec \textit{et al.}~\cite{kervadec2019boundary} introduced a boundary loss that uses a distance metric on the space of boundaries for medical image segmentation to enhance the segmentation accuracy near the boundary.
Karimi \textit{et al.}~\cite{karimi2019reducing} proposed a hausdorff distance loss to reduce the hausdorff distance directly to improve the similarity between the predicted mask and ground truth.
Shit \textit{et al.}~\cite{shit2021cldice} incorporated morphological operation into loss function calculation of thin objects, and proposed a clDice loss for accurate boundary segmentation.
Jia \textit{et al.}~\cite{jia2019hd} introduced a 2D auxiliary boundary detection branch that shares the encoder with the 3D segmentation branch and achieved superior performance on MRI prostate segmentation.
Despite the improved performance, the boundary-involved loss function-based methods~\cite{kervadec2019boundary,karimi2019reducing,shit2021cldice} are time-consuming during training since these boundary-involved loss functions are computing-unfriendly, $i.e.$, the distance map of each training image mask should be computed during each iteration.
Whereas the previous auxiliary boundary task-based method~\cite{jia2019hd} does not fully utilize the extracted boundary information since the auxiliary boundary branch will be abandoned at inference time.
In our previous work~\cite{hu_boundary-aware_2020}, we adopted a 3D auxiliary boundary decoder and introduced the skip connections from the boundary decoder to the segmentation decoder to boost the kidney tumor segmentation performance.

In this paper, we propose a boundary-aware network (BA-Net) based on our previous work~\cite{hu_boundary-aware_2020} for kidney, renal tumor, renal artery, and renal vein segmentation. 
The BA-Net is an encoder-decoder structure~\cite{ronneberger_u-net_2015}.
To force the model to pay more attention to the error-prone boundaries, we introduce an auxiliary boundary decoder to detect kidney-related structures' ambiguous boundaries. 
Compared to our previous work in~\cite{hu_boundary-aware_2020}, both the boundary and segmentation decoders are supervised at each scale of the decoder in this paper to further improve the model's robustness to varying sizes of target organs.
Also, we modified the feature fusion mechanism and used the detected boundary probability maps as attention maps on the corresponding segmentation features instead of directly concatenating the boundary feature maps with segmentation feature maps.
We have evaluated the proposed BA-Net model on the KiPA Challenge training dataset using 4-fold cross-validation and achieved a Dice score of 96.59$\%$ for kidney segmentation, 90.74$\%$ for renal tumor segmentation, 87.75$\%$ for renal artery segmentation, and 83.53$\%$ for renal vein segmentation.

\section{Dataset}
The kidney-related structures segmentation dataset published by KiPA challenge\footnote{\url{https://kipa22.grand-challenge.org/}} was used for this study.
The KiPA dataset contains 130 CT scans with voxel-level annotations of 4 kidney structures, including the kidney, renal tumor, renal artery and renal vein.
Among all the scans, 70 scans are provided as training cases, 30 scans are used for open test evaluation, and the left 30 scans are withheld for close test evaluation. 
Only the voxel-level annotations of training cases are publicly available, while the annotations of open test cases and close test cases cannot be accessed.

\begin{figure}[t]
  \centering
  \includegraphics[width=1\textwidth]{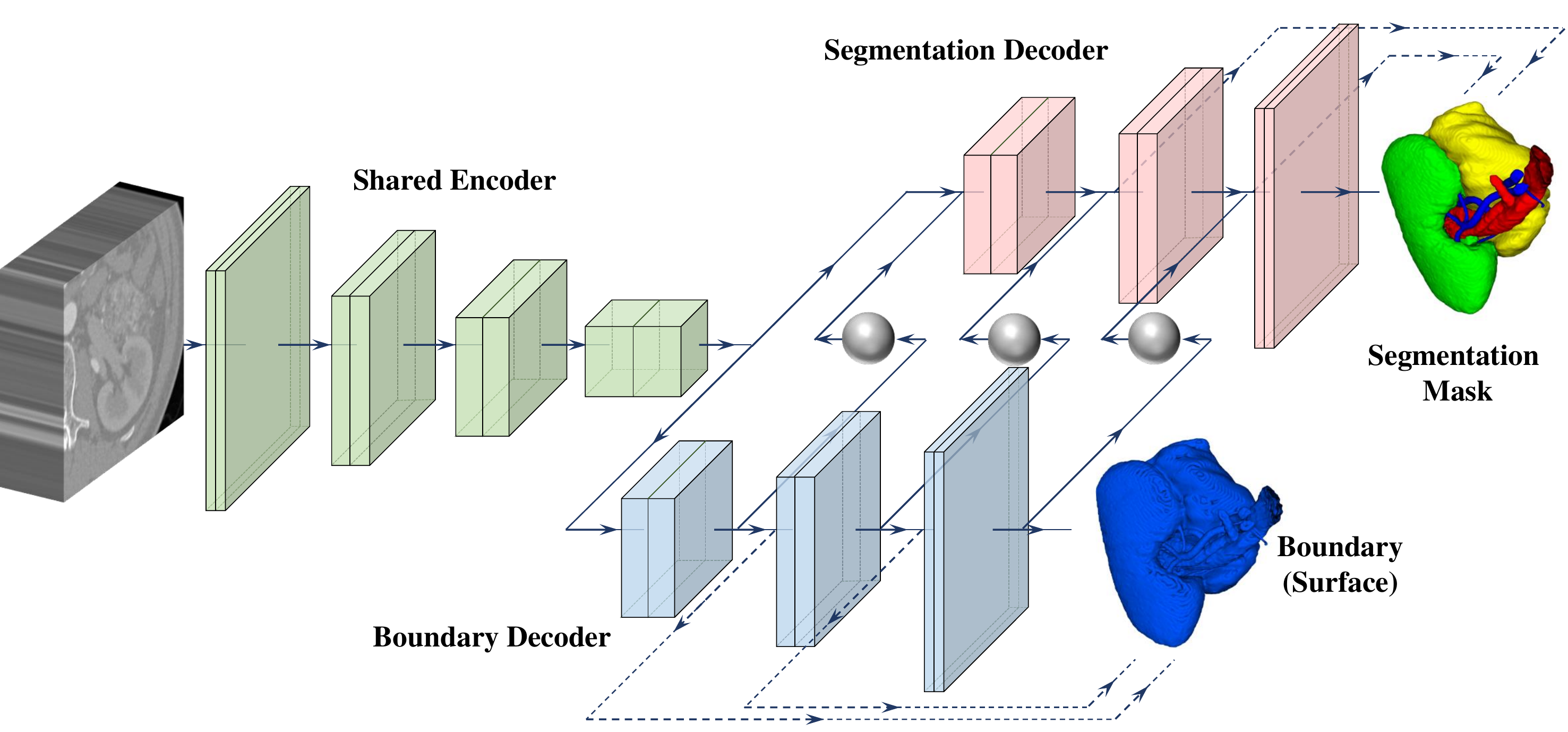}
  \caption{Illustration of BA-Net for kidney-related structures segmentation. The green, blue, and pink layers represent convolutional layers of the encoder, boundary decoder, and segmentation decoder respectively. The gray nodes represent the soft-max operations. The dashed lines in both decoders represent deep supervision. The skip connections from the shared encoder to both decoders are omitted for simplicity.}
  \label{fig:overview}
\end{figure}

\section{Method}
The proposed BA-Net contains a shared encoder, a boundary decoder, and a segmentation decoder, as shown in~\figurename{~\ref{fig:overview}}.
The encoder extracts features from the input images. 
Then the boundary decoder performs boundary detection using the extracted features. 
The segmentation decoder takes the extracted features as input and outputs segmentation maps of target structures, wherein the boundary probability maps produced by the boundary decoder are adopted as attention maps to enhance the segmentation features.
\subsection{Shared Encoder}
Since nnUNet~\cite{isensee_nnu-net_2021} performs well on kidney tumor segmentation~\cite{heller2021state}, we adopt it as the backbone of our BA-Net.
Therefore, the shared encoder design is the same as nnUNet, which is generated from the statistics of the training data.
It contains $N$ encoder blocks ($N=5$ for kidney structures segmentation task), each of which is composed of 2 convolutional layers.
Each convolutional layer is followed by instance normalization and the LeakyReLU activation.
The down-sample operations are performed using strided convolutions, followed after each encoder block.
In the encoder, the number of convolutional filters is set to 32 in the first layer, then doubled in each next block, and finally fixed with 320 when it becomes larger than 256.

\subsection{Boundary Decoder}
Symmetrically, the boundary decoder contains $N$ decoder blocks to upsample the feature map extracted by the encoder and gradually refine it to generate the boundary map.
The transposed convolution with a stride of 2 is used to improve the resolution of input feature maps.
The upsampled feature map is concatenated with the low-level feature map extracted from the corresponding encoder block and then fed to the decoder block.
The output feature map $f_i^b$ of each decoder block is processed by a convolutional layer and a soft-max layer to generate the boundary probability map $p_i^b$ at each scale $i$, where $i=1,2,\cdots,N-1$.

\subsection{Segmentation Decoder}
The structure of the segmentation decoder is similar to the boundary decoder, except for an additional step to enhance the segmentation feature map at each scale.
In each block of the segmentation decoder, the upsampled feature map $U(f_{i-1}^s)$ is enhanced with the corresponding boundary probability map $p_i^b$, shown as follows

\begin{equation}
\begin{aligned}
\widehat{U(f_{i-1}^s)} = (1 + p_i^b) * U(f_{i-1}^s)
\end{aligned}
\end{equation}
where $\widehat{U(f_{i-1}^s)}$ is the enhanced feature map, and $U(\cdot)$ represents upsample operation.

The enhanced feature map is concatenated with the feature map extracted from the corresponding encoder block before being further processed by the segmentation decoder block.
Similar to the boundary decoder, the output feature map $f_i^s$ of each segmentation decoder block is also processed by a convolutional layer and a soft-max layer to generate the segmentation probability map $p_i^s$ at each scale $i$.

\subsection{Training and Inference}
The combination of Dice loss and cross-entropy loss is adopted as the objective for both the boundary detection and segmentation tasks.
The joint loss at each scale can be calculated as

\begin{equation}
\begin{aligned}
\mathcal{L}_i=& 1-\frac{2 \sum_{v=1}^{V} p_{iv} y_{iv}}{\sum_{v=1}^{V}\left(p_{iv}+y_{iv}+\epsilon\right)} \\
&-\sum_{v=1}^{V}\left(y_{iv} \log p_{iv}+\left(1-y_{iv}\right) \log \left(1-p_{iv}\right)\right)
\end{aligned}
\end{equation}
where $p_{iv}$ and $y_{iv}$ denote the prediction and ground truth of the $v$-th voxel in the output of the $i$-th decoder block, $V$ represents the number of voxels, and $\epsilon$ is a smooth factor to avoid dividing by 0.
For the segmentation task, $p_{i}$ is the segmentation probability map $p_i^s$, and $y_i$ is the downsampled segmentation ground truth $y_i^s$. For the boundary detection task, $p_{i}$ is the boundary probability map $p_i^b$, and $y_i$ is the boundary $y_i^b$ of the segmentation ground truth $y_i^s$.

We use deep supervision for both decoders to ensure the boundary can be detected at each scale to enhance the segmentation feature map and improve the model's robustness to targets' sizes.
Totally, the loss is defined as

\begin{equation}
\begin{aligned}
\mathcal{L} = \sum_{i=1}^{N-1}\omega_i(\mathcal{L}_i^{s} + \mathcal{L}_i^{b})
\end{aligned}
\end{equation}
where $\omega_i$ is a weighting vector that enables higher resolution output to contribute more to the total loss~\cite{isensee_nnu-net_2021}, $\mathcal{L}_i^{s}$ is the segmentation loss, and $\mathcal{L}_i^{b}$ is the boundary detection loss.

During inference, given a test image, the shared encoder extracts features at each scale, and then each boundary decoder block processes these features and produces a boundary probability map.
Based on the extracted features and the boundary probability maps, the segmentation decoder is only required to output the segmentation result of the last decoder block.

\section{Experiments and Results}
\subsection{Implementation Details and Evaluation Metric}
The CT scans were resampled to a uniform voxel size of $3mm\times0.6mm\times0.6mm$.
The Hounsfield Unit (HU) values of CT scans were clipped to the range of $[918, 1396]$ according to the data statistics, and then subtracted the mean and divided by the standard deviation.
We applied data augmentation techniques, including random cropping, random rotation, random scaling, random flipping, random Gaussian noise addition, and elastic deformation to generate augmented input volumes with a size of $112\times128\times128$ voxels.
Limited to the GPU memory, the batch size was set to $2$.
The SGD algorithm was used as the optimizer. The initial learning rate $lr_0$ was set to 0.01 and decayed according to $lr = lr_0\times(1-t/T)^{0.9}$, where $t$ is the current epoch and $T$ is the maximum epoch.
$T$ was set to 200 for model selection, and 1000 for online evaluation.
The whole framework was implemented in PyTorch and trained using an NVIDIA 2080Ti.

The segmentation performance was measured by the Dice Similarity Coefficient (DSC) and Hausdorff Distance (HD).
All experiments were performed using 4-fold cross-validation with connected components-based postprocessing.

\subsection{Results}

\begin{table}[]
\centering
\caption{Performance (DSC \%, HD voxel) of BA-Net and six competing methods in kidney structures segmentation on KiPA \textbf{training} dataset. The models were trained for \textbf{200} Epochs. The best results are highlighted with \textbf{bold}.}
\label{tab:results}
\setlength\tabcolsep{1.5pt}
\renewcommand{\arraystretch}{1.2}
\begin{tabular}{l|cc|cc|cc|cc|cc}
\hline \hline
\multirow{2}{*}{Methods} & \multicolumn{2}{c|}{Kidney} & \multicolumn{2}{c|}{Renal Tumor} & \multicolumn{2}{c|}{Renal Artery} & \multicolumn{2}{c|}{Renal Vein} & \multicolumn{2}{c}{\textbf{Average}} \\
\cline{2-11}
                         & DSC $\uparrow$          & HD $\downarrow$          & DSC $\uparrow$          & HD $\downarrow$            & DSC $\uparrow$          & HD $\downarrow$             & DSC $\uparrow$          & HD $\downarrow$            & \textbf{DSC} $\uparrow$          & \textbf{HD} $\downarrow$          \\
\hline
nnUNet~\cite{isensee_nnu-net_2021}   & 96.30 & 2.44 & 88.50 & 6.35 & 87.35 & 1.60 & 82.53 & 5.63 & 88.67 & 4.01          \\
\hline
clDice~\cite{shit2021cldice}       & 95.83 & 3.65 & 87.64 & 9.06 & 86.14 & 3.50 & 81.59 & 7.75 & 87.80 & 5.99           \\
\hline
BD Loss~\cite{kervadec2019boundary}& 96.19 & 1.95 & 88.34 & 7.55 & 86.90 & 1.95 & 82.01 & 5.77 & 88.36 & 4.31            \\
\hline
HD Loss~\cite{karimi2019reducing}& 96.12 & 2.58 & 88.48 & 7.22 & 86.40 & 2.37 & 82.10 & 5.76 & 88.28 & 4.48           \\
\hline
ResUNet~\cite{isensee_nnu-net_2021}& 96.33 & 2.70 & \textbf{89.92} & 6.70 & 87.15 & 2.35 & 82.63 & 6.24 & 89.01 & 4.50           \\
\hline
BA-Net*~\cite{hu_boundary-aware_2020}& 96.39 & 1.87 & 89.29 & 5.79 & 87.23 & 1.90 & \textbf{83.55} & 6.07 & 89.12 & 3.91           \\
\hline
BA-Net (Ours)                   & \textbf{96.47} & \textbf{1.59} & 89.74 & \textbf{4.78} & \textbf{87.59} & \textbf{1.25} & 83.17 & \textbf{4.10} & \textbf{89.24} & \textbf{2.93}           \\
\hline \hline
\end{tabular}
\end{table}

\begin{table}[]
\centering
\caption{Performance (DSC \%, HD voxel) of BA-Net and three competing methods in kidney structures segmentation on KiPA \textbf{training} dataset. The models were trained for \textbf{1000} Epochs. The best results are highlighted with \textbf{bold}. `RB-Ensemble' represent the performance of the averaged result of ResUNet and BA-Net.}
\label{tab:results-1000}
\setlength\tabcolsep{1.5pt}
\renewcommand{\arraystretch}{1.2}
\begin{tabular}{l|cc|cc|cc|cc|cc}
\hline \hline
\multirow{2}{*}{Methods} & \multicolumn{2}{c|}{Kidney} & \multicolumn{2}{c|}{Renal Tumor} & \multicolumn{2}{c|}{Renal Artery} & \multicolumn{2}{c|}{Renal Vein} & \multicolumn{2}{c}{\textbf{Average}} \\
\cline{2-11}
                         & DSC $\uparrow$          & HD $\downarrow$          & DSC $\uparrow$          & HD $\downarrow$            & DSC $\uparrow$          & HD $\downarrow$             & DSC $\uparrow$          & HD $\downarrow$            & \textbf{DSC} $\uparrow$          & \textbf{HD} $\downarrow$          \\
\hline
nnUNet~\cite{isensee_nnu-net_2021}   & 96.40 & 1.89 & 90.23 & 6.97 & 87.30 & 1.63 & 82.91 & 5.14 & 89.21 & 3.91           \\
\hline
ResUNet~\cite{isensee_nnu-net_2021}& 96.48 & 1.79 & 90.62 & 5.41 & 87.33 & 1.51 & 82.81 & 5.37 & 89.31 & 3.52            \\
\hline
BA-Net                   & 96.54 & 1.81 & 90.22 & \textbf{4.61} & 87.36 & \textbf{1.18} & 83.02 & 4.98 & 89.28 & \textbf{3.14}            \\
\hline
RB-Ensemble                   & \textbf{96.59} & \textbf{1.79} & \textbf{90.74} & 5.36 & \textbf{87.48} & 1.44 & \textbf{83.21} & \textbf{4.01} & \textbf{89.51} & 3.15            \\
\hline \hline
\end{tabular}
\end{table}

\begin{table}[]
\centering
\caption{Performance (DSC \%, HD $mm$) of BA-Net and four competing methods in kidney structures segmentation on KiPA \textbf{open test} dataset. The best results are highlighted with \textbf{bold}. `RB-Ensemble' represent the performance of the averaged result of ResUNet and BA-Net.}
\label{tab:openset}
\setlength\tabcolsep{0.8pt}
\renewcommand{\arraystretch}{1.2}
\begin{tabular}{l|cc|cc|cc|cc|cc}
\hline \hline
\multirow{2}{*}{Methods} & \multicolumn{2}{c|}{Kidney} & \multicolumn{2}{c|}{Renal Tumor} & \multicolumn{2}{c|}{Renal Artery} & \multicolumn{2}{c|}{Renal Vein} & \multicolumn{2}{c}{\textbf{Average}} \\
\cline{2-11}
                         & DSC $\uparrow$          & HD $\downarrow$          & DSC $\uparrow$          & HD $\downarrow$            & DSC $\uparrow$          & HD $\downarrow$             & DSC $\uparrow$          & HD $\downarrow$            & \textbf{DSC} $\uparrow$          & \textbf{HD} $\downarrow$          \\
\hline
DenseBiasNet~\cite{he2020dense}    & 94.60 & 23.89 & 79.30 & 27.97 & 84.50 & 26.67 & 76.10 & 34.60 & 83.63 & 28.28 \\
\hline
MNet~\cite{dong2022mnet}    & 90.60 & 44.03 & 65.10 & 61.05 & 78.20 & 47.79 & 73.50 & 42.60 & 76.85 & 48.87 \\
\hline
3D U-Net~\cite{cciccek20163d}  & 91.70 & 18.44 & 66.60 & 24.02 & 71.90 & 22.17 & 60.90 & 22.26 & 72.78 & 21.72 \\
\hline
\hline
BA-Net                   & 95.87 & \textbf{16.30} & 89.23 & 12.07 & 87.53 & \textbf{14.87} & \textbf{85.32} & 12.53 & 89.49 & 13.94             \\
\hline
RB-Ensemble                   & \textbf{95.91} & 16.61 & \textbf{89.64} & \textbf{11.52} & \textbf{87.78} & 15.18 & 85.30 & \textbf{12.27} & \textbf{89.66} & \textbf{13.89}            \\
\hline \hline
\end{tabular}
\end{table}

We compared our BA-Net with 
(1) two SOTA medical image segmentation methods: nnUNet and ResUNet~\cite{isensee_nnu-net_2021};
(2) three boundary-involved loss function-based methods: clDice~\cite{shit2021cldice}, BD Loss~\cite{kervadec2019boundary}, and HD Loss~\cite{karimi2019reducing};
and (3) our previous boundary-involved segmentation approach: BA-Net*~\cite{hu_boundary-aware_2020},
as shown in~\tablename{~\ref{tab:results}}.
It shows that the overall performance of our BA-Net is not only better than nnUNet and ResUNet, but also superior to boundary-involved loss function-based methods.
Also, the BA-Net outperforms our previous BA-Net*.
It demonstrates that the boundary probability attention map can enhance the segmentation features effectively.

Since more training iterations under strong data augmentation can improve the segmentation performance, we sufficiently retrained nnUNet, ResUNet, and our BA-Net for 1000 epochs.
\tablename{~\ref{tab:results-1000}} shows the performance of these methods.
It shows that the performance of the model can be further improved when trained in more iterations.
Still, our BA-Net achieves better performance than nnUNet and is comparable with ResUNet.
Therefore, we averaged the results produced by ResUNet and our BA-Net, denoted as RB-Ensemble.
It reveals from~\tablename{~\ref{tab:results-1000}} that the averaged results achieved the best DSC, but slightly worse HD than our BA-Net.

We also evaluated the performance of our BA-Net and the ensemble result on the open test set of the KiPA challenge dataset.
\tablename{~\ref{tab:openset}} shows the performance of these two methods and three approaches designed for kidney parsing.
It demonstrates that our BA-Net achieves much better performance than the previous kidney structure segmentation methods, confirming the effectiveness of our BA-Net.
\section{Conclusion}
In this paper, we propose a BA-Net for kidney-related structure segmentation.
It is composed of a shared encoder for feature extraction, a boundary decoder for boundary detection, and a segmentation decoder for target organ segmentation.
At each scale of the boundary decoder and the segmentation decoder, the boundary and segmentation mask at the corresponding scale are used to supervise the predicted probability map.
The predicted boundary map at each decoder block is used as an attention map to enhance the segmentation features.
Experimental results on the KiPA challenge dataset demonstrate the effectiveness of our BA-Net.

\bibliographystyle{splncs04}
\bibliography{reference}

\begin{thebibliography}{10}
\providecommand{\url}[1]{\texttt{#1}}
\providecommand{\urlprefix}{URL }
\providecommand{\doi}[1]{https://doi.org/#1}

\bibitem{cciccek20163d}
{\c{C}}i{\c{c}}ek, {\"O}., Abdulkadir, A., Lienkamp, S.S., Brox, T.,
  Ronneberger, O.: 3d u-net: learning dense volumetric segmentation from sparse
  annotation. In: International conference on medical image computing and
  computer-assisted intervention. pp. 424--432. Springer (2016)

\bibitem{dong2022mnet}
Dong, Z., He, Y., Qi, X., Chen, Y., Shu, H., Coatrieux, J.L., Yang, G., Li, S.:
  Mnet: Rethinking 2d/3d networks for anisotropic medical image segmentation.
  arXiv preprint arXiv:2205.04846  (2022)

\bibitem{he2020dense}
He, Y., Yang, G., Yang, J., Chen, Y., Kong, Y., Wu, J., Tang, L., Zhu, X.,
  Dillenseger, J.L., Shao, P., et~al.: Dense biased networks with deep priori
  anatomy and hard region adaptation: Semi-supervised learning for fine renal
  artery segmentation. Medical image analysis  \textbf{63},  101722 (2020)

\bibitem{he2021meta}
He, Y., Yang, G., Yang, J., Ge, R., Kong, Y., Zhu, X., Zhang, S., Shao, P.,
  Shu, H., Dillenseger, J.L., et~al.: Meta grayscale adaptive network for 3d
  integrated renal structures segmentation. Medical Image Analysis
  \textbf{71},  102055 (2021)

\bibitem{heller2021state}
Heller, N., Isensee, F., Maier-Hein, K.H., Hou, X., Xie, C., Li, F., Nan, Y.,
  Mu, G., Lin, Z., Han, M., et~al.: The state of the art in kidney and kidney
  tumor segmentation in contrast-enhanced ct imaging: Results of the kits19
  challenge. Medical image analysis  \textbf{67},  101821 (2021)

\bibitem{hu_boundary-aware_2020}
Hu, S., Zhang, J., Xia, Y.: Boundary-aware network for kidney tumor
  segmentation. In: Liu, M., Yan, P., Lian, C., Cao, X. (eds.) {MICCAL} 2020 -
  {MLMI}. pp. 189--198. Lecture Notes in Computer Science, Springer
  International Publishing. \doi{10.1007/978-3-030-59861-7_20}

\bibitem{huang2022adwunet}
Huang, Z., Wang, Z., zhikai yang, Gu, L.: Adwu-net: Adaptive depth and width
  u-net for medical image segmentation by differentiable neural architecture
  search. In: Medical Imaging with Deep Learning (2022),
  \url{https://openreview.net/forum?id=kF-d1SKWJpS}

\bibitem{isensee_nnu-net_2021}
Isensee, F., Jaeger, P.F., Kohl, S.A.A., Petersen, J., Maier-Hein, K.H.:
  {nnU}-net: a self-configuring method for deep learning-based biomedical image
  segmentation  \textbf{18}(2),  203--211. \doi{10.1038/s41592-020-01008-z},
  \url{https://www.nature.com/articles/s41592-020-01008-z}

\bibitem{jia2019hd}
Jia, H., Song, Y., Huang, H., Cai, W., Xia, Y.: Hd-net: hybrid discriminative
  network for prostate segmentation in mr images. In: International Conference
  on Medical Image Computing and Computer-Assisted Intervention. pp. 110--118.
  Springer (2019)

\bibitem{karimi2019reducing}
Karimi, D., Salcudean, S.E.: Reducing the hausdorff distance in medical image
  segmentation with convolutional neural networks. IEEE Transactions on medical
  imaging  \textbf{39}(2),  499--513 (2019)

\bibitem{kervadec2019boundary}
Kervadec, H., Bouchtiba, J., Desrosiers, C., Granger, E., Dolz, J., Ayed, I.B.:
  Boundary loss for highly unbalanced segmentation. In: International
  conference on medical imaging with deep learning. pp. 285--296. PMLR (2019)

\bibitem{milletari2016v}
Milletari, F., Navab, N., Ahmadi, S.A.: V-net: Fully convolutional neural
  networks for volumetric medical image segmentation. In: 2016 fourth
  international conference on 3D vision (3DV). pp. 565--571. IEEE (2016)

\bibitem{Peng_2022_CVPR}
Peng, C., Myronenko, A., Hatamizadeh, A., Nath, V., Siddiquee, M.M.R., He, Y.,
  Xu, D., Chellappa, R., Yang, D.: Hypersegnas: Bridging one-shot neural
  architecture search with 3d medical image segmentation using hypernet. In:
  Proceedings of the IEEE/CVF Conference on Computer Vision and Pattern
  Recognition (CVPR). pp. 20741--20751 (June 2022)

\bibitem{ronneberger_u-net_2015}
Ronneberger, O., Fischer, P., Brox, T.: U-net: Convolutional networks for
  biomedical image segmentation. In: Navab, N., Hornegger, J., Wells, W.M.,
  Frangi, A.F. (eds.) {MICCAI} 2015. pp. 234--241. Lecture Notes in Computer
  Science, Springer International Publishing.
  \doi{10.1007/978-3-319-24574-4_28}

\bibitem{shao2011laparoscopic}
Shao, P., Qin, C., Yin, C., Meng, X., Ju, X., Li, J., Lv, Q., Zhang, W., Xu,
  Z.: Laparoscopic partial nephrectomy with segmental renal artery clamping:
  technique and clinical outcomes. European urology  \textbf{59}(5),  849--855
  (2011)

\bibitem{shao2012precise}
Shao, P., Tang, L., Li, P., Xu, Y., Qin, C., Cao, Q., Ju, X., Meng, X., Lv, Q.,
  Li, J., et~al.: Precise segmental renal artery clamping under the guidance of
  dual-source computed tomography angiography during laparoscopic partial
  nephrectomy. European urology  \textbf{62}(6),  1001--1008 (2012)

\bibitem{shit2021cldice}
Shit, S., Paetzold, J.C., Sekuboyina, A., Ezhov, I., Unger, A., Zhylka, A.,
  Pluim, J.P., Bauer, U., Menze, B.H.: cldice-a novel topology-preserving loss
  function for tubular structure segmentation. In: Proceedings of the IEEE/CVF
  Conference on Computer Vision and Pattern Recognition. pp. 16560--16569
  (2021)

\end{thebibliography}

\end{document}